\documentclass[showpacs,nofootinbib]{revtex4}
\usepackage{latexsym}
\usepackage{amsfonts}
\usepackage{amsbsy}
\usepackage{mathrsfs}

\def\ut#1{\rlap{\lower1ex\hbox{$\sim$}}#1{}}

\begin{document}

\title{Husain-Kuchar model as a constrained BF theory}

\date{\today}

\author{Merced Montesinos}\email{merced@fis.cinvestav.mx}

\author{Mercedes Vel\'azquez}
\email{mquesada@fis.cinvestav.mx}

\affiliation{Centre de Physique Th\'eorique, CNRS Luminy, F-13288 Marseille,
France}

\affiliation{Departamento de F\'{\i}sica, Cinvestav, Instituto Polit\'ecnico
Nacional 2508, San Pedro Zacatenco, 07360, Gustavo A. Madero, Ciudad de
M\'exico, M\'exico.}

\begin{abstract}
The Husain-Kuchar theory is a four-dimensional background-independent model
that has long been viewed as a useful model for addressing several conceptual
and technical problems appearing in the quantization of general relativity
mainly in the loop quantum gravity approach. The model was defined at
Lagrangian level in terms of a $su(2)$-valued connection one-form $A$ coupled
through its curvature to a $su(2)$-valued one-form field $e$. We address here
the problem of writing a Lagrangian formulation for the Husain-Kuchar model as
a constrained BF theory motivated by the fact that spin foam models for
quantum gravity are related to action principles of the BF type. The
Lagrangian action principle for the Husain-Kuchar model reported here differs
from a previous one found by Barbero {\it et al} in that this description
involves a single constrained BF theory rather than two interacting BF
theories. It is, essentially, the Plebanski action with the condition on the
trace of the Lagrange multipliers removed. Moreover, it can be stated that the
relationship between our BF-like action and the original one for the
Husain-Kuchar model is the same relationship that exists between the Plebanski
action and the self-dual Palatini action for complex general relativity, first
because the solution to the constraint on the two-forms $\Sigma^i$ coming from
the BF-like action leads to the Husain-Kuchar action, and second because the
Hamiltonian analysis of the Husain-Kuchar model is straightforward starting
from the BF-like action principle.
\end{abstract}


\maketitle

It is very well-know that general relativity, expressed in terms of Ashtekar
variables \cite{1Ashtekar}, and the Husain-Kuchar model \cite{hkmodel} are
very close to each other. Their similarities are usually appreciated by
employing the Hamiltonian form of their corresponding action principles. In
particular, these theories have the same phase space variables but the
Hamiltonian constraint of general relativity is missing in the Husain-Kuchar
model, which is defined by the action principle \cite{hkmodel}
\begin{eqnarray}\label{luminy}
S[e,A] = \int_{\mathscr{M}^4} \left [ \varepsilon_{ijk} e^j \wedge e^k
\wedge F^i [A] \right ],
\end{eqnarray}
where $F^i [A] = d A^i + \varepsilon^i\,_{jk} A^j \wedge A^k$ is the curvature
of the $su(2)$-valued connection 1-form $A=A^i J_i$, $e=e^i J_i$ is a
$su(2)$-valued one-form field, and $J^i$ are the generators of $su(2)$ and
satisfy $[J_i , J_j]= \varepsilon^k\,_{ij} J_k$.

Perhaps one of the best ways to understand the relationship between the
self-dual Palatini action \cite{1Samuel,1Jacobson} and the action principle
(\ref{luminy}) is to look at the self-dual Palatini action as a constrained
Husain-Kuchar model obtained by adding the constraint $e^i \wedge F_i [A]$ to
the action ({\ref{luminy}) with a Lagrange multiplier 1-form $\lambda$
\cite{bar}. On the other hand, the Husain-Kuchar model can also be described
by an action principle involving $e^i$, $A^i$, and a scalar field $\phi$
enlarging in this way the phase space but introducing an additional scalar
constraint which allows it to preserve the three local degrees of freedom of
the model \cite{bar2}. Even though these works are very interesting and
valuable because they shed light on the subject, the current research in
quantum gravity (spin foam models) suggest to look for BF-like formulations
for the Husain-Kuchar model with the hope that these can help to understand
the role of the Hamiltonian constraint of general relativity at the quantum
level.

In last line of thought and at the classical level, there exist a previous
work by Barbero {\it et al} in which the Husain-Kuchar model is described as
two interacting unconstrained $SO(3)$ $BF$ theories \cite{bvmodel}.

In this paper, in opposition to the viewpoint adopted in Ref. \cite{bvmodel},
we study the Husain-Kuchar model as a single constrained BF theory. In
particular, we want to see the differences and similarities between this
theory and general relativity when both of them are formulated in the BF
Lagrangian framework and we want to distinguish one from the other theory by
the way the constraints on the two-forms fields are imposed. Following this
line of thought, it is quite natural to look at the various ways general
relativity is formulated as a constrained BF theory
\cite{jmp1977,penrose,capo,cqg1999,cqg1999b,rob,cqgl2001} (see also Refs.
\onlinecite{vlad,vlad2}) and, in particular, to the first of these
formulations which was given by Plebanski \cite{jmp1977} and, inspired by it,
try to find a formulation for the Husain-Kuchar model.

Throughout the paper, Greek indices $\mu,\nu, \ldots =0,1,2,3$ are spacetime
indices that label the points of the spacetime $\mathscr{M}^4$, Latin
lowercase indices $a,b,c,\ldots =1,2,3$ denote space indices. When the
canonical analysis is performed ${\mathscr{M}}^4$ is assumed to have the form
${\mathscr{M}}^4 = \mathbb{R}\times \Sigma$ with $\Sigma$ compact and without
a boundary (to avoid boundary terms) and $(x^{\mu})=(x^0,x^a)$ with $x^0$ and
$x^a$ labeling the points along $\mathbb{R}$ and $\Sigma$, respectively.

Let us first recall the Plebanski's formulation for general relativity as a
constrained BF theory, which is based on the action
\begin{eqnarray}\label{plebaction}
S[\Sigma, A, C,\rho] &=& \int_{\mathscr{M}^4} \left [ \Sigma_i \wedge F^i [A] -
\frac12 C_{ij} \Sigma^i \wedge \Sigma^j + \rho \left ( C^i\,_i - \Lambda \right ) \right ],
\end{eqnarray}
where $A= A^i J_i$ is a connection one-form valued in the complexification of
$su(2)$ and $F=F^i[A] J_i$ with $F^i[A] = d A^i + \frac12 \varepsilon^i\,_{jk}
A^j \wedge A^k$ is its curvature, $\Sigma= \Sigma^i J_i$ is a two-form valued
in the complexication of $su(2)$, $J^i$ are the generators of the $su(2)$ Lie
algebra and satisfy the commutation relations $[J_i , J_j]=
\varepsilon^k\,_{ij} J_k$, the Lagrange multipliers $C_{ij}$ form a symmetric
matrix $(C_{ij})=(C_{ji})$, $\rho$ is a 4-form field, and $\Lambda$ is the
cosmological constant.

The variation of the action (\ref{plebaction}) with respect to all the
independent fields involved yields the equations of motion
\begin{eqnarray}
\begin{array}{llr}
\delta A^i: & D \Sigma^i = 0, & (3\times 4 =12 \,\,
\mbox{equations}), \\
\delta \Sigma^i: & F^i [A] =  C^i\,_j \Sigma^j, & (3\times 6 =18
\,\, \mbox{equations}), \\
\delta C_{ij}: & -\Sigma^i \wedge \Sigma^j + 2 \rho \delta^{ij} =
0, & (6 \,\,\mbox{equations}), \\
\delta \rho: & C^i\,_i - \Lambda =  0, & (1 \,\, \mbox{equation}).
\end{array}
\end{eqnarray}
Alternatively, Plebanski's action (\ref{plebaction}) acquires the equivalent
form
\begin{eqnarray}\label{pleba2}
S \left [\Sigma, A, M \right ] &=& \int_{\mathscr{M}^4} \left [
\Sigma_i \wedge F^i [A] - \frac12 M_{ij}  \Sigma^i \wedge \Sigma^j
- \frac{  \Lambda}{6} \Sigma^i \wedge \Sigma_i  \right ],
\end{eqnarray}
which involves only $35$ independent variables because the matrix $(M_{ij})$
is now traceless from the very beginning. The equations of motion obtained
from the action (\ref{pleba2}) are
\begin{eqnarray}\label{emotion}
\begin{array}{llr}
\delta A^i : & D \Sigma^i = 0, & (3\times 4 =12 \,\,
\mbox{eqns}), \\
\delta \Sigma^i: & F^i [A] =  M^i\,_j \Sigma^j + \frac{\Lambda}{3} \Sigma^i, & (3\times 6 =18
\,\, \mbox{eqns}), \\
\delta M_{ij}: & \Sigma^i \wedge \Sigma^j - \frac13 \delta^{ij} \Sigma^k \wedge \Sigma_k =
0, & (5 \,\,\mbox{eqns}).
\end{array}
\end{eqnarray}
The five equations for the $3\times 6=18$ variables $\Sigma^i\,_{\mu\nu}$
given in (\ref{emotion}) define a $13$-dimensional manifold $\mathcal{M}^{13}$
embedded in the $18$-dimensional space whose points are coordinatized by the
variables $\Sigma^i\,_{\mu\nu}$. This way of looking at $\mathcal{M}^{13}$ is
directly related to the Lagrangian action principle (\ref{pleba2}) but there
exists another, equivalent form, of looking at $\mathcal{M}^{13}$ which is
useful to go straightforwardly to its Hamiltonian formulation. This second
viewpoint follows from the fact that the points of
 $\mathcal{M}^{13}$ can be put in one-to-one correspondence with the points of a $13$-dimensional
manifold $\mathcal{N}^{13}$ defined by the nine equations
\begin{eqnarray}\label{plebsol}
\Sigma^i\,_{0a} = - \frac14 \ut{\,\varepsilon}\,_{abc} \left [  N^b {\widetilde \pi}^{ic} + \ut{N}
\varepsilon^i\,_{jk} {\widetilde \pi}^{jb} {\widetilde \pi}^{kc}  \right ],
\end{eqnarray}
on the 22 coordinates $\Sigma^i\,_{0a}$, ${\widetilde \pi}^{ia}$, $N^a$, and
$\ut{N}$ that label the points of a 22-dimensional manifold. In Eq.
(\ref{plebsol}), ${\widetilde \pi}^{ia} = {\widetilde \varepsilon}^{abc}
\Sigma^i_{bc}$ with ${\widetilde \varepsilon}^{abc}= {\widetilde
\varepsilon}^{0abc}$ the Levi-Civita tensor density. More precisely, the five
equations for $\Sigma^i$ given in (\ref{emotion}) can be seen as five linear
equations for the nine variables $\Sigma^i\,_{0a}$ and thus the general
solution of the system of equations must involve four arbitrary functions,
which are the densitizied lapse $\ut{N}$ and shift $N^a$ that appear as
Lagrange multipliers associated to the Hamiltonian and vector constraints,
respectively (see Refs. \cite{japan,cdjm}).

Let us go now to the Husain-Kuchar model. From the already well-known fact
that at the Hamiltonian level this model has not a Hamiltonian constraint, it
follows that the analog of Eq. (\ref{plebsol}) for the Husain-Kuchar model is
\begin{eqnarray}\label{hksol}
\Sigma^i\,_{0a} = - \frac14 \ut{\,\varepsilon}\,_{abc} N^b {\widetilde \pi}^{ic}.
\end{eqnarray}
In analogy to what (\ref{plebsol}) does for general relativity, the nine
equations (\ref{hksol}) define a $12$-dimensional manifold $\mathcal{H}^{12}$
embedded in a $21$-dimensional manifold whose points are coordinatized by
$\Sigma^i\,_{0a}$, ${\widetilde \pi}^{ia}$, and $N^a$. This viewpoint will be
used below to go to the Hamiltonian formulation of the theory. However, to
build the Lagrangian principle we are looking for, it is required to find an
equivalent $12$-dimensional manifold $\mathcal{F}^{12}$ suitable for the
Lagrangian formulation in the sense that its definition be covariant and
involves the $\Sigma^i$ only. Alternatively, a careful analysis of the way
that the terms involving the lapse $\ut{N}$ cancel to each other when the
solution (\ref{plebsol}) for general relativity is inserted back into the
equation $\Sigma^i \wedge \Sigma^j - \frac13 \delta^{ij} \Sigma^k \wedge
\Sigma_k=0$ leads us to the covariant equation that (\ref{hksol}) must satisfy
for the Husain-Kuchar model. This equation is
\begin{eqnarray}\label{hkbasic}
\Sigma^i \wedge \Sigma^j =0.
\end{eqnarray}
In fact, the six equations for the $3\times 6=18$ variables
$\Sigma^i\,_{\mu\nu}$ given in (\ref{hkbasic}) define the $12$-dimensional
manifold $\mathcal{F}^{12}$ embedded in the $18$-dimensional space whose
points are coordinatized by the $\Sigma^i\,_{\mu\nu}$. The points of
$\mathcal{F}^{12}$ can be put in one-to-one correspondence with the points of
the $12$-dimensional manifold $\mathcal{H}^{12}$. Equivalently, if the
equation (\ref{hksol}) is inserted back into (\ref{hkbasic}), these equations
are automatically satisfied.

Therefore, the constrained BF action for the Husain-Kuchar model we are
proposing is, essentially, the one obtained from the Plebanski action
(\ref{plebaction}) by removing the condition $C^i\,_i - \Lambda=0$ on the
Lagrange multipliers $C_{ij}$. To be precise, the action is given by
\begin{eqnarray}\label{hk}
S[\Sigma, A, C] &=& \int_{\mathscr{M}^4}
\left [ \Sigma_i \wedge F^i [A] - \frac12 C_{ij} \Sigma^i \wedge \Sigma^j  \right ],
\end{eqnarray}
where there are six independent components in the symmetric matrix ($C_{ij})$.
The variation of the action (\ref{hk}) gives the equations of motion
\begin{eqnarray}\label{hkem}
\begin{array}{llr}
\delta A^i: & D \Sigma^i = 0, & (3\times 4 =12 \,\,
\mbox{equations}), \\
\delta \Sigma^i: & F^i [A] =  C^i\,_j \Sigma^j, & (3\times 6 =18
\,\, \mbox{equations}), \\
\delta C_{ij}: & \Sigma^i \wedge \Sigma^j =
0, & (6 \,\,\mbox{equations}),
\end{array}
\end{eqnarray}
With the use of (\ref{hksol}), the Hamiltonian analysis of the action
(\ref{hk}) is straightforward. In fact, if (\ref{hksol}) is inserted back into
the action (\ref{hk}), it acquires the Hamiltonian form
\begin{eqnarray}
S[A^i_a, {\widetilde \pi}^a_i, \lambda^i, N^a] = \int_{\mathscr{M}^4} d^4 x
\left [ {\dot A}^i_a {\widetilde \pi}^a_i -
\lambda^i \widetilde{{\mathcal G}}_i - N^a {\widetilde V}_a \right ],
\end{eqnarray}
where
\begin{eqnarray}
\widetilde{{\mathcal G}}_i &=& D_a {\widetilde \pi}_i^a \approx 0, \nonumber\\
{\widetilde V}_a &=& F^i\,_{ab} {\widetilde \pi}^b_i \approx 0,
\end{eqnarray}
are the Gauss and vector first-class constraints corresponding to the
Husain-Kuchar model \cite{hkmodel}. Therefore, actions (\ref{luminy}) and
(\ref{hk}) are dual to each other in the sense that both describe the same
(classical) physics but they involve different fields.

The action principle (\ref{hk}) is of the form $S[B^M,A^N] = \int_{{\mathscr
M}^4} B^M \wedge F_N [A]$, where $F= F^M[A] X_M$ with $F^M[A]= d A^M + \frac12
C^M\,_{NL} A^N \wedge A^L$ is the curvature of a ${\frak g}$-valued connection
one-form $A= A^M X_M$, $B= B^M X_M$ is also a ${\frak g}$-valued $2$-form,
Latin capital indices $M,N, L, \dots = 1, \ldots, \rm{dim} ({\frak g})$ are
Lie algebra indices that are raised and lowered with respect to $k_{MN}:= k
(X_M , X_N)$, which is a scalar product among the generators with respect to
an inner product $k$ in the Lie algebra ${\frak g}$. The inner product is
assumed to be invariant under the action of the group which means that the
structure constants $C^L\,_{MN}$ are totally antisymmetric. The generators
$X_M$ satisfy $[X_M,X_N]= C^L\,_{MN} X_L$. This action principle for the
particular Lie algebras $SO(3,1)$ and $SO(4)$ was reported in Ref.
\onlinecite{cqg2003}. However, its relationship with the Husain-Kuchar model
was not set there. More recently, it is mentioned in Ref. \onlinecite{kirill}
that (\ref{hk}) is general relativity with the Hamiltonian constraint removed.
Nevertheless, it is not mentioned there to which of all Hamiltonian
formulations of general relativity (\ref{hk}) is related nor how this
relationship can be achieved. Moreover, the relationship of (\ref{hk}) to the
Husain-Kuchar model is also not made there.

Thus, we have shown the equivalence between the action principle (\ref{hk})
and the Husain-Kuchar model (\ref{luminy}) by performing the canonical
analysis of the action (\ref{hk}) and showing that the result so obtained is
the one that follows from the canonical analysis of the action (\ref{luminy}).
The fact that the Hamiltonian analysis of the Husain-Kuchar model can be
carried out so easily is one of the advantages of the action principle
(\ref{hk}) over the action (\ref{luminy}). Notice that this property of the
action (\ref{hk}) is similar to the one of the Plebanski action
(\ref{plebaction}) in the sense that also the Hamiltonian analysis of the
Plebanski action leads immediately to the Hamiltonian description of general
relativity in terms of Ashtekar variables \cite{japan,cdjm}, in opposition to
what is usually done and that consists in performing the Dirac analysis to the
self-dual Palatini action or performing a complex canonical transformation
from the triad and the extrinsic curvature \cite{1Samuel,1Jacobson,1Ashtekar}.

Up to here, the main results are the action principle (\ref{hk}) and its
Hamiltonian analysis. It is possible to go further. In the same sense that the
Plebanski action (\ref{plebaction}) leads to the self-dual Palatini action for
general relativity introduced by Samuel and Jacobson and Smolin by solving the
constraint $\Sigma^i \wedge \Sigma^j -\frac13 \delta^{ij} \Sigma^k \wedge
\Sigma_k=0$ in the third line of (\ref{emotion}) by using the so-called
reality conditions which allows it to write $\Sigma^i$ in terms of a real
tetrad as $\Sigma^i= i e^0 \wedge e^i - \frac12 \varepsilon^i\,_{jk} \wedge
e^j \wedge e^k$ \cite{cdjm}, it is also possible to obtain action
(\ref{luminy}) from action (\ref{hk}) by solving for $\Sigma^i$ the constraint
(\ref{hkbasic}) of the BF description for the Husain-Kuchar model as (modulo a
constant factor) the product of the 1-forms $e^i$
\begin{eqnarray}\label{keykey}
\Sigma^i = \varepsilon^i\,_{jk} e^j \wedge e^k,
\end{eqnarray}
which amounts to parameterize the surface $\mathcal{F}^{12}$ in terms of the
$3\times 4=12$ variables $e^i_{\mu}$. It is clear that Eq. (\ref{hkbasic}) is
automatically satisfied by inserting in that equation this expression for
$\Sigma^i$. On the other hand, by plugging (\ref{keykey}) into the action
(\ref{hk}), it becomes (\ref{luminy}), as expected. Therefore, we have shown
that the Lagrangian action principle for the Husain-Kuchar model can also be
obtained from (\ref{hk}) by solving the constraint (\ref{hkbasic}).

Once we know that (\ref{hkbasic}) implies (\ref{keykey}), it is natural to ask
if the action principle (\ref{hk}) can be obtained from an BF action principle
by adding to it the equation (\ref{keykey}) with a Lagrange multiplier
$\lambda^i$ that is a 2-form field. By doing this, we get
\begin{eqnarray}\label{rb}
S[\Sigma,A,e,\lambda] &=& \int_{\mathscr{M}^4} \left[ \Sigma_i \wedge F^i [A] -
\lambda_i \wedge \left ( \Sigma^i - \varepsilon^i\,_{jk} e^j \wedge e^k \right ) \right].
\end{eqnarray}
The variation of the action (\ref{rb}) with respect to the independent fields
leads to the equations of motion
\begin{eqnarray}\label{emrb}
&& \delta A^i : D \Sigma^i =0, \nonumber\\
&& \delta \Sigma^i: F^i [A]= \lambda^i, \nonumber\\
&& \delta \lambda^i: \Sigma^i = \varepsilon^i\,_{jk} e^j \wedge e^k, \nonumber\\
&& \delta e^i: \varepsilon_{ijk} \lambda^j \wedge e^k = 0.
\end{eqnarray}
From these it is clear that the action (\ref{luminy}) can be independently
obtained either by plugging back the expression for $\lambda^i$ or the
expression for $\Sigma^i$ given in the second and third rows of (\ref{emrb})
into the action (\ref{rb}). Nevertheless, it is also possible to solve for
$\lambda^i$ by using the last equation in (\ref{emrb}). In fact, the system of
twelve linear equations for eighteen unknowns $\lambda^i_{\alpha\beta}$ is of
rank twelve generically, which means that six out of the Lagrange multipliers
$\lambda^i_{\alpha\beta}$ can be chosen as free parameters. Alternatively,
such an equation defines a $18$-dimensional manifold embedded in the space
whose points are coordinatized by $\lambda^i_{\alpha\beta}$ and $e^i\,_{\mu}$.
By putting all this together, the solution for $\lambda^i$ acquires the form
\begin{eqnarray}\label{ya}
\lambda^i = C^i\,_j \left ( \varepsilon^j\,_{mn} e^m \wedge e^n \right ),
\end{eqnarray}
which depends on eighteen independent variables: six variables encoded in the
symmetric matrix $(C_{ij})$ plus twelve involved in $e^i_{\mu}$, as it should
be. Therefore, the last equation in (\ref{emrb}) is automatically satisfied by
plugging (\ref{ya}) into that equation while the remaining equations become
\begin{eqnarray}\label{yaya}
&& D \Sigma^i =0, \nonumber\\
&& F^i [A]= C^i\,_j \left ( \varepsilon^j\,_{mn} e^m \wedge e^n \right ), \nonumber\\
&& \Sigma^i = \varepsilon^i\,_{jk} e^j \wedge e^k,
\end{eqnarray}
which, using what we already have shown, namely, that the equation in the
third row implies (\ref{hkbasic}), the system of equations (\ref{yaya})
becomes that given in (\ref{hkem}), which can be obtained from (\ref{hk}).
This way of getting (\ref{hk}) is a direct application of the so-called {\it
parent action method} to the Husain-Kuchar action (\ref{luminy}) \cite{papa}.

In summary, the close relationship at the Hamiltonian level between the
Husain-Kuchar model and general relativity expressed in terms of Ashtekar
variables can be clearly appreciated when both theories are expressed at
Lagrangian level as constrained BF theories. The action for the Husain-Kuchar
model is the one given in (\ref{hk}) while general relativity is described by
the Plebanski action (\ref{plebaction}). The difference between the two
theories lies in the fact that the Plebanski action (\ref{plebaction})
involves one condition more than the BF-like action principle (\ref{hk}) for
the Husain-Kuchar model: the condition on the trace of the matrix $(C_{ij})$.
At the Hamiltonian level this condition appears as one constraint more on the
phase space variables: the Hamiltonian constraint, which is missing in the
Hamiltonian formulation of the Husain-Kuchar model, i.e., it is also possible
to say that the action for general relativity (\ref{plebaction}) is the
Husain-Kuchar action supplemented with a constraint.


We conclude the paper by making some remarks and pointing out possible
implications of our result:

(1) Note that if a cosmological term $\Sigma^i \wedge \Sigma_i$ is added to
the action (\ref{hk}), the constraint (\ref{hkbasic}) and its solution
(\ref{hksol}) still hold.

(2) Even though the current analysis was performed in the self-dual case, we
would not expect any qualitative change between the self-dual and the real
formulations, for instance, the ones obtained from the action principles
considered in Refs. \onlinecite{cqgl2001} and \onlinecite{epr} by removing the
condition of the trace on the Lagrange multipliers $\phi_{IJKL}$ involved in
such formulations.

(3) Due to the fact the Husain-Kuchar model has not a Hamiltonian constraint,
it might be possible that a quantization of the theory based in the action
(\ref{hk}) reported in this paper (or in its Euclidean version), or in the
ones suggested in item (2) using the tools of the spin foam models can help to
better understand the role of the Hamiltonian constraint operator (through its
absence) in the spin foam approach to the quantization of the gravitational
field, by comparing these (hypothetical) spin foam models and a spin foam
model for general relativity. If this could be done, such a result would be
also very useful to better understand the interplay between the canonical and
covariant quantizations for gravity involved in the loop quantum gravity and
spin foam models approaches, respectively \cite{spinfoam,lqg}(for a spin foam
quantization of the Husain-Kuchar model see Ref. \onlinecite{bpibe}). More
precisely, our results strongly suggest that it is possible to make a slightly
modification to the spin foam model for gravity considered in Ref. \cite{epr}
in order to build a new one compatible with the quantum version of the
constraint (\ref{hkbasic}). This is linked to the item (2). The discretized
version of the constraint seems to be a condition on the 4-simplex saying that
it has zero volume. This is currently under investigation as well as an
Euclidean version for the quantum Husain-Kuchar model and quantum gravity
\cite{joaquim} inspired by the model considered in Ref. \cite{lattice}.

Warm thanks to Alejandro Perez and Carlo Rovelli for very fruitful discussions
on the subject of this paper. We thank Kirill Krasnov for drawing our
attention to Ref. \cite{kirill}. This work was supported in part by CONACYT,
Mexico, Grant Numbers 56159-F and 79629 (sabbatical term). M. Vel\'azquez also
acknowledges the financial support from CONACYT. We thank the {\it Centre de
Physique Th\'eorique} at Luminy, Marseille for all support and facilities
provided for the realization of the sabbatical term of M. Montesinos and the
research term of M. Vel\'azquez.

\end{document}